\documentclass[fleqn,usenatbib]{mnras} 

\usepackage[T1]{fontenc}
\usepackage{ae,aecompl}

\def\spose#1{\hbox to 0pt{#1\hss}}
\def\lta{\mathrel{\spose{\lower 3pt\hbox{$\mathchar"218$}}
     \raise 2.0pt\hbox{$\mathchar"13C$}}}
\def\gta{\mathrel{\spose{\lower 3pt\hbox{$\mathchar"218$}}
     \raise 2.0pt\hbox{$\mathchar"13E$}}}
 
\def\1p{\phantom{0}}
\def\2p{\phantom{00}}
\def\3p{\phantom{000}}
\def\4p{\phantom{0000}}

\usepackage{graphics,epsfig}
\usepackage{graphicx}

\title[Activity-evolution of the Hyades K giants]
{Magnetic activity and evolution of the four Hyades K giants}

\author[K.-P. Schr\"oder, M. Mittag, D. Jack, A. Rodr\'iguez 
Jim\'enez, J.H.M.M. Schmitt]
{K.-P. Schr\"oder$^{1,3}$\thanks{email: kps@astro.ugto.mx},
M. Mittag$^{2}$, D. Jack$^{1,2}$, A. Rodr\'iguez Jim\'enez$^{1}$, 
\newauthor 
J.H.M.M. Schmitt$^{2}$ \\
$^{1}$Depto. Astronomia, Universidad de Guanajuato, 
A.P. 144, Guanajuato, GTO, C.P. 36000, Mexico 
\\
$^{2}$Hamburger Sternwarte, Universit\"at Hamburg, Gojenbergsweg 112, 
D-21029 Hamburg, Germany \\
$^{3}$ Sterrewacht Leiden, Nils Bohrweg 2, NL-2333CA Leiden, Netherlands
}

\begin{document}

\pagerange{\pageref{firstpage}--\pageref{lastpage}} 
\pubyear{2019}

\maketitle

\label{firstpage}

\begin{abstract}
We determine the exact physical parameters of the four Hyades cluster K giants,
using their parallaxes and atmospheric modeling of our red-channel TIGRE
high-resolution spectra. 
Performing a comparison with well-tested evolutionary tracks, we derive exact 
masses and evolutionary stages. At an age of 588 ($\pm 60$) Myrs and 
with a metallicity of Z=0.03 (consistent with the spectroscopic abundances), 
we find HD~27371 and HD~28307, the two less bright K giants, at the onset 
of central helium-burning, entering their blue loops with a mass of 
2.62 $M_{\odot}$, while the slightly brighter stars HD~28305 and HD~27697 are 
already exiting their blue loop. Their more advanced evolution suggests a 
higher mass of 2.75 $M_{\odot}$.

Notably, this pairing coincides with the different activity levels, 
which we find for these four stars from chromospheric 
activity monitoring with TIGRE and archival Mount Wilson data as well
as from ROSAT coronal detections: 
The two less evolved K giants are the far more active pair, and we
confidently confirm their rotation with periods of about 142 days. This 
work therefore provides some first, direct evidence of magnetic 
braking during the 130~Myrs lasting phase of central helium-burning, 
similar to what has long been known to occur to cool main-sequence stars. 
 \end{abstract}


\section{Introduction}

After the discovery of chromospheric Ca II H\&K line emission as a
possible activity indicator by \cite{eberhard1913}, 
it was mainly Olin C. Wilson and his collaborators at the Mount Wilson 
Observatory, who carried out much of
the early work on chromospheric activity. The Ca II H\&K line 
emission was observed for all kinds of stars \citep{duncan1991}
across the HR diagramme (HRD), 
and a spectroscopic activity monitoring programme of solar-like 
stars resulted in the discovery of stellar activity cycles similar 
to the solar 11 year Schwabe cycle \citep{baliunas1995}.

Since the early 1960ies, this research benefited from the introduction 
of an easily measurable quantity, the so-called S-index, and a specific 
four-channel narrow-band spectrophotometer designed by O.C. Wilson for making 
instantaneous S-index measurements (see \cite{vaughan1978} with more details 
given below).  As a consequence, the magnetic activity observed for the Sun
could finally be juxtaposed to the activity observed for stars and thus the
Sun could be put in its proper stellar perspective.

O.C. Wilson's historic work on chromospheric Ca II H\&K emission of giant
stars is fundamental (e.g. \cite{wilson1957}), since a number of brighter 
giants had been  included in his chromospheric activity monitoring programme,
which started in the 1960ies.
However, since most of the cooler giants do not show coronal
X-ray emission \citep{linshaisch1979}, doubt was cast 
on their activity, until it became understood (see, e.g, \cite{ayres1997}), 
that magnetic activity does continue into the most evolved stages of stellar 
evolution. The more recent findings of direct evidence of magnetic fields in 
giant stellar photospheres leave no doubt on this issue
(\cite{hubrig1994}, \cite{konst2013}). 

Unfortunately, this line of research of the Wilson group was never published 
in a refereed 
publication.  In fact, cool giants are a mixed group of stars with 
very different masses and different evolutionary states, in addition, 
there are observational and other complications (see \cite{schro2018} 
for a recent discussion).
Most K giants, however, are simpler. The large majority is in 
the relatively stable phase of central helium 
burning, and as such are intermediate to giants on the red giant branch
(RGB) and asymptotic giant branch (AGB).
 
As a consequence, the Hyades K giants are of a particular interest for an 
understanding of how magnetic activity evolves with the star.
Ever since \cite{sku1972} demonstrated a relation between increasing age and 
decreasing activity of a star, the importance of magnetic braking of stellar
rotation has been recognized.  In the case of main sequence stars (see, e.g., \cite{schro2013} 
and references given therein), in particular from measurements of the chromospheric
Ca II H\&K emission there is convincing observational evidence for the action
of magnetic braking;
in that work we also find that activity on the main
sequence appears to decline with the age {\it relative} to the star's
main sequence life-time, since both, magnetic breaking and evolution time-scales
depend on the stellar mass in a similar way. 

Also, from X-ray detections across 
the HRD we know that recent Hertzsprung gap crossers are 
very active again, and that the K clump giants of the solar neighbourhood 
still have moderate coronal emission -  their range of X-ray surface fluxes 
is comparable to that of the Sun \citep{huensch1996a}. With magnetic 
activity also found on the AGB (see Duncan et al. 1991,
Schr\"oder et al. 2018),
the main question here is, whether we observe
a down- or an upturn of activity during central helium burning. Such 
empirical evidence is crucial to understand the evolution of the 
internal angular momentum and how it shapes dynamo action as
well as the interplay with magnetic braking. 

Therefore, the four bright and nearby Hyades K giants are an obvious starting 
point for such research. Already more than 35 years ago, \cite{baliunas1983} 
combined observations of chromospheric emission by means of the 
S-index and UV emission lines observed by IUE as well as coronal X-ray 
detections from the {\it Einstein Observatory}, to come to ask the important 
question: Why and how can these four giants of the same age and almost
the same mass and structure be so different in their activity levels? 

Based on his chromospheric activity monitoring measurements, this
 question occurred to  O.C. Wilson already in 1972
  (private communication through D. Reimers to one of us (KPS)).
When our robotic 1.2 m telescope TIGRE with its 
high-resolution spectrograph HEROS started its operations at Guanajuato in 
2014 \citep{schmitt2014}, all four Hyades K giants were included
in TIGRE's chromospheric activity monitoring programme.

Consequently, this paper looks at the exact evolution states of these
four K giants (section 2) and at their activity (section 3),
by briefly reviewing the X-ray observations
and analyzing in depth the chromospheric activity monitoring of
the Mount Wilson group and by ourselves with TIGRE.

\section{Evolutionary states: a small but crucial difference}

Before turning to a discussion of the activity properties of the Hyades 
giants, 
we begin with an assessment of the individual evolutionary states of these 
objects. The Hyades cluster is not very rich in stars; for example, in the
recent study of Hyades membership based on Gaia parallaxes
\cite{lodieu2019} find 710 members within a distance of 30~pc from
the cluster center, with 85 members being located in the actual cluster 
core, and
the same authors identify 8 brown dwarfs and verify their Hyades membership.
Yet, the Hyades also contain a sufficient number of evolved stars. Again 
based on Gaia parallaxes, \cite{salaris2018} find at least 8 white dwarfs as 
Hyades cluster members, which as the more massive stars have already 
passed the red giant stage.

The four prominent K-type Hyades giants ($\epsilon$ Tau, $\delta$ Tau, 
$\gamma$ Tau, $\theta ^1$ Tau) have been known for a long time.  
This surprisingly large number is based on the speed of stellar evolution: 
The stars in question have masses of about 
2.7 $M_{\odot}$ (see our discussion below), and their 
central helium-burning lifetime of about 130~million years accounts 
for as much as 20\% of the total stellar lifetime.  
In the case of the Sun, for example, central Helium-burning accounts only
for 1\% of its lifetime, and therefore such low-mass K giants are found in
much smaller fractions (say one per hundred main sequence stars).   
Consequently, the number of 
K clump giants can be a lot smaller in clusters, which differ from 
the Hyades in turn-off mass and hence age.
This little detail makes the Hyades K giants an
excellent test bed for a study of how stellar evolution evolves
during central helium burning.

\subsection{Assessment of the exact physical parameters}

The two most important stellar parameters, distinguishing the Hyades 
K giants from one another, are luminosity $L$ and effective temperature 
$T_{\rm eff}$. Since the individual parallaxes of the Hyades giants
are now known,  the main remaining uncertainty comes from the
values adopted for the bolometric correction and the solar bolometric
magnitude.  In Table 1 we list the parallaxes,
luminosities and all other stellar quantities relevant for our study.

For the luminosities as given in Table 1, we use 
BC~=~0.50 for all four K giants, since their colours are very 
similar, and M$_{\rm Bol \odot}$= 4.74, consistent with 
long-standing calibrations like the one given by \cite{SK1982}. 
At least this choice does not affect the relative 
differences between these giants, which matter most for this study. Somewhat 
smaller bolometric corrections favoured by more recent compilations,
which suggest a BC around 0.4 for the Hyades K giants
(e.g. \cite{flower1996}), would result in slightly smaller 
masses, i.e., less by up to 2.5\% or 0.07 $M_{\odot}$ for all four
giants, and an age up to 10\% (60 Myrs) larger than the values given below,
but there would be no change in the relative differences.

To derive effective temperatures from our high s/n
(up to 200) $R=21000$ TIGRE/HEROS spectra, we use 
the spectra analysis tool iSpec \citep{blanco2014},
working on the orange-red part of the spectrum, from which we 
excluded small spectral regions with line blends, to avoid confusion.
Regions contaminated with telluric lines have also been excluded 
from the analysis, which is based on a comparison with 
a library of synthetic spectra (ATLAS9 of \citet{kurucz}) and 
the line list of the Vienna Atomic Line Data Base (VALD, 
\citet{vald}), employing the SPECTRUM code. As a reference to 
the solar abundances, we use \citet{asplund2009}. 

To obtain the most reliable results, we follow the recommendation
of \cite{blanco2019} to keep as many parameters 
fixed as possible, which reduces the errors of the parameters to be derived. 
In our case, we derive the mass from matching evolution tracks, and 
with a preliminary value for $T_{\rm eff}$ we then calculate $\log{g}$, 
which we find close to 2.5 in all four stars, and so use it as a fixed value 
in the automatic iSpec analysis. 
We estimate the remaining uncertainty in $T_{\rm eff}$ 
to be under 100 K or 0.008 dex (see error bars in Fig. 1).
The abundances as obtained by this method by us
lie in the range given by a vast literature (see, e.g., 
\cite{perry1998}, \cite{ramy2019} and citations therein), 
  and fall around [Fe/H]=0.22. 

\begin{table*}
	\caption{Astrophysical quantities and activity indicators of the four 
		Hyades K giants. Parallax values are courtesy to 
		GAIA DR2, obtained from VizieR} 
		of the Strasbourg astronomical data centre. 
	\begin{small}
		\begin{tabular}{l | l |l | l | l | l | l | l | l } 
			\hline
	star & $\pi$ & $\log{L}$ & $\log{T_{\rm eff}}$ & $M$ & $R$ & $<S>$ 
	& $L_X$ & $F_x$  \\ 
	& [mas] & [$L_{\odot}$] & [K] & [$M_{\odot}$] & [$R_{\odot}$] & 
	& [$10^{29} erg s^{-1}$] & [$10^{4} erg~cm^{-2} s^{-1}$]   \\ 
			\hline
    HD 28305/$\epsilon$ Tau  & 20.31 & 2.07 &  3.681 & 2.75 &  15.8 & 0.129  & 0.15 & 0.10 \\
	HD 27697/$\delta$ Tau    & 19.06 & 2.03 &  3.683 & 2.75 &  14.9 & 0.133  & 0.54 & 0.40 \\
	HD 27371/$\gamma$ Tau    & 22.62 & 1.93 &  3.681 & 2.62 &  13.4 & 0.178  &11.60 &10.45 \\
	HD 28307/$\theta^1$ Tau  & 21.42 & 1.90 &  3.683 & 2.62 &  12.8 & 0.166  &18.43 &18.16 \\
			\hline		
		\end{tabular}
	\end{small}
	\label{Tab:SynPar}
\end{table*}

\begin{figure}
	\centering
	\begin{tabular}{c}
\includegraphics[width=0.85\linewidth,angle=-90]{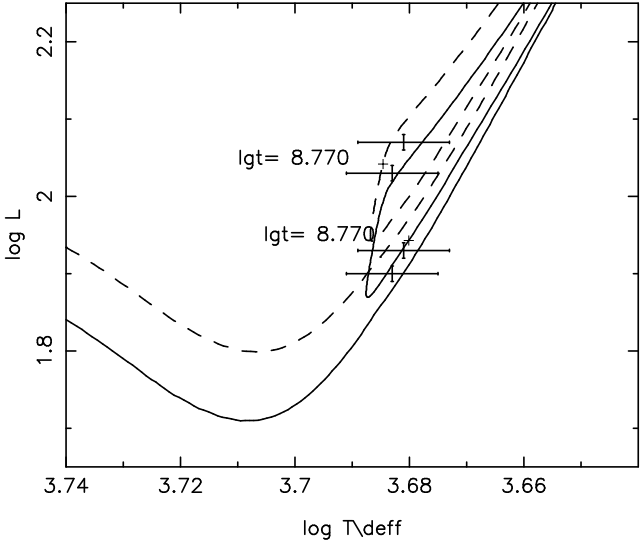}
	\end{tabular}
	\caption{Evolution models for a metallicity of Z=0.03 (close to
          [Fe/H]=+0.2) and with stellar masses of 2.62 (solid line)
          and 2.75 $M_{\odot}$ (- -) fit 
	all four Hyades K giants at their HRD positions for an age of 
	588 Myrs: HD 27371 and HD 28307  at the onset of central 
	helium-burning, which lasts 130 Myrs and forms the blue loop, 
	while the slightly brighter pair of
	HD 28305 and HD 27697 is already exiting the blue loop.}
	\label{fig_track}
\end{figure}

\subsection{Derivation of the exact states of evolution}

The HRD positions resulting above need to be compared to suitable evolution 
models, which we computed with the Cambridge (UK) Eggleton code in its
updated version. 
Significant improvements were made with respect to 
opacities and equation of state as described by \cite{pols97} and  \cite{pols98};
in these papers a detailed description of our procedures is given, here we
provide only a summary. 
While the code is using a classical mixing-length-theory approach to convection,
it differs from others in solving an additional equation
to optimize the spread of its height points and concentrate them 
in layers of steep gradients like burning shells. This concept makes the code
economic with respect to CPU time and robust, as it works well with only 
200 height points.

For central hydrogen burning models for main sequence stars, like 
all evolution codes, the luminosity at any given mass depends 
on the used equation of state and 
on the choice of helium abundance. For cool stars with large
convective envelopes, their radii and effective temperatures depend on
the parameterization of the convective length scale in terms of the pressure
scaleheight, $l_c=\alpha H_P$.  As described in the first of the above 
papers, we use a best-choice of $\alpha=2.0$, where $\alpha=2.0$.
In addition, for stars with masses larger than about
1.5 $M_{\odot}$, the prescription of convective core overshoot,
which empirically and indiscriminately includes all kind of extra
mixing beyond the Schwarzschildt boundary, plays an
important role from the later central hydrogen burning onwards.

The second of the above papers demonstrates an excellent agreement of these 
models of the code (as used here) with eclipsing binaries of well-known
physical parameters. These models also agree very well with those of the 
classical Geneva code of \cite{meyn1993}. Apparently, slightly different choices of 
the helium abundance Y (adopting a slightly different $\Delta{Y}/\Delta{Z}$)
compensated for small differences in the equation of state. Our models use
Y~=~0.28 for a nearly solar metalicity of Z=0.02, and Y=0.30 for Z=0.03
(as then used here for modeling the moderately iron-rich Hyades giants, 
and for the models shown in Fig.~\ref{fig_track}).
Later, based on another type of evolutionary code, models were
published by \cite{pietr2004} and \cite{pietr2006}, which also agree very well
with the resulting physical parameters of our models. 

The amount of core overshooting prescribed on the medium-mass, central hydrogen
burning of a stellar model has an ever increasing effect on the mass of the 
resulting helium core in the evolving star. 
Consequently, the brightness of the resulting
red giant provides -- especially during the central helium burning (blue loop)
phase -- a very sensitive test of this issue, whenever the respective stellar
mass is well known. This idea was carefully employed by \cite{schro1997}, using
giants in eclipsing ($\zeta$ Aur type) binaries with masses, luminosities,
radii and effective temperatures well known from observations. Based on this
approach, the resulting overshoot parameterization (used here with the very 
same code) empirically accounts for all extra mixing beyond the convective, 
hydrogen-burning core (apart from genuine overshooting, this can be, e.g. 
the average meridional mixing by rotation). For the masses of the Hyades 
K giants (see below), our models employ a nominal overshoot length of 
$l_{ov}=0.24 H_P$, which is slowly
rising to $0.3 H_P$ for larger masses, see Fig.~10 in  \cite{schro1997}.   
For further technical detail on the evolution code and its parameterization used
here, we like to refer the reader again to \cite{pols97} and
\cite{pols98}. 

Assuming that all four Hyades giants  have the same age, 
they then must have slightly different masses to show somewhat different
advances of their stellar evolution. For this assessment, we compared a variety of
evolutionary tracks to the HRD positions obtained above. Where these tracks, to within the
observational uncertainties, can be met by different stages of evolution,
we gave preference to the slowest phase 
(such as the blue loop, which marks the stable central helium burning), 
since the probability for
finding a star at point in the HRD is simply larger. The same argument makes a fast phase,
e.g., the very swift ascent on the RGB or the climbing of the AGB, a per se
unlikely choice. The other strongly discriminating condition is, as pointed
out above, that all final models must not only match the observed HRD positions,
but also have the same age.

In this fashion and using the physical quantities given in Table 1, we find
masses of  (i) 2.62 $M_{\odot}$ for the less bright
pair, which has just come down the RGB to start central helium burning and
turned off the main sequence only 10 Myrs ago, and
(ii) of  2.75 $M_{\odot}$ for the brighter pair, which is about to finish 
central helium burning, which according to our models lasts 130 Myrs.
This is long enough to tell us about evolutionary
changes in the level of the K giant activity, as we will demonstrate in the
next section.
We reemphasize our assumption that all four giants have the same age, which
implies that we attribute  their evolutionary
differences, i.e., start and end of central helium burning, 
solely to the mass difference of 0.13 $M_{\odot}$, which causes the two more
luminous and more massive giants to be more evolved than the less massive
giant pair.

Since all other possible matches with evolutionary tracks would 
set either the one or the other pair of giants outside the blue-loop 
segment -- that is: not one, but always two stars have a less likely, fast
evolutionary state -- the solution presented here in Fig.~\ref{fig_track}
is significantly more probable than any other formally possible solution, 
since for the four giants
taken together, the probabilities of matching any individual star multiply with
each other. In consequence, this approach of ruling out other matching choices
involving faster episodes in evolution, is much stronger for a set of stars like the four
K giants studied here than it is for a single star.

The Hyades cluster age of our models shown in Fig.~\ref{fig_track}, is 588 Myrs,
in good agreement with \cite{gossage2018} (in particular with their
model for a rotation of $\Omega/\Omega_c=0.6$) and references
given therein.
A long-standing literature age of the Hyades is 625 Myrs (\cite{perry1998}).
As mentioned above, the uncertainty in luminosity of about 10\% due to
a possibly lower BC of 0.4 would give us alternative models with slightly
smaller masses and, consequently, of an age of up to 648 Myrs. Hence, the
age agreement lies reasonably well within this and other uncertainties.
Furthermore, we should note that our models shown in  Fig.~\ref{fig_track}
suggest a mass of 2.60 $M_{\odot}$ at the turn-off point.

\section{Activity of the Hyades K giants}

\subsection{Coronal activity: X-ray emission}

In the X-ray range \cite{stern1981}, 
using the {\it Einstein Observatory}, obtained
the first detections of X-ray emission from the Hyades giants
and measured X-ray luminosities of 10$^{29.4}$ erg/s,  10$^{28.9}$ erg/s, 
and 10$^{30.0}$ erg/s for the stars $\gamma$~Tau, $\delta$~Tau, and 
$\theta^1$~Tau, respectively, while the star  $\epsilon$~Tau was outside
the field of view and thus remained unobserved by the
{\it Einstein Observatory}.   

Naturally, the Hyades region was scanned
in the context of the ROSAT all-sky survey (RASS) and in their 
RASS study of the Hyades region \cite{stern1995} report the X-ray detections
of all four Hyades giants.  Using the most recent RASS catalog by
\cite{boller2016} we identify the RASS sources
2RXS~J041947.5+153739 with $\gamma$~Tau,
2RXS~J042252.4+173148 with $\delta$~Tau,
2RXS~J042836.6+191036 with $\epsilon$~Tau and
2RXS~J042834.7+155721 with $\theta^1$~Tau, respectively, which were
observed with count rates of 0.44 $\pm$ 0.04 cts/s, 0.03 $\pm$ 0.01 cts/s, 
0.02 $\pm$ 0.01 cts/s, and 0.81 $\pm$ 0.04 cts/s respectively.

Hence, the dichotomy between $\gamma$ Tau and $\theta^1$ Tau 
(HD 27371 and HD 28307) on the one hand, and $\delta$ Tau and $\epsilon$ Tau 
(HD 27697 and HD 28305) on the other hand is immediately apparent in the
observed RASS rates, with clear detections for the former (active) pair, and
almost marginal detections for the latter pair of Hyades K giants.  However, 
later pointed ROSAT observations of $\delta$ Tau (in ROSAT sequence RP200442) 
with a count rate of 0.0250 $\pm$ 0.0012 cts/s and of $\epsilon$ Tau
(in ROSAT sequence RP200576) with a count rate of 0.0101 $\pm$ 0.0007 cts/s
provided clear confirmations of all RASS detections.  

Using these count rates
we can compute X-ray fluxes, X-ray luminosities and X-ray surface fluxes, 
using the count-flux-conversion by \cite{schmitt1995}, and thus arrive at the 
numbers quoted in Table 1.   It is important to keep in mind
that detailed spectral information is not available from the ROSAT data,
the flux conversion is therefore fraught with considerable uncertainty,
which we estimate to be on the order of 50~\%. And we need to keep in mind
that, while active stars are variable, these X-ray detections are based
on only a small number of visits.

Nevertheless, a bifurcation is clearly seen into two
active and two inactive K giants, very much like (as shown below) is
the result of chromospheric activity monitoring, even though X-ray data
do not yield any significant ranking order within each of these two pairs.

\begin{figure}
	\centering
	\begin{tabular}{c}
	\includegraphics[width=0.90\linewidth,angle=0]{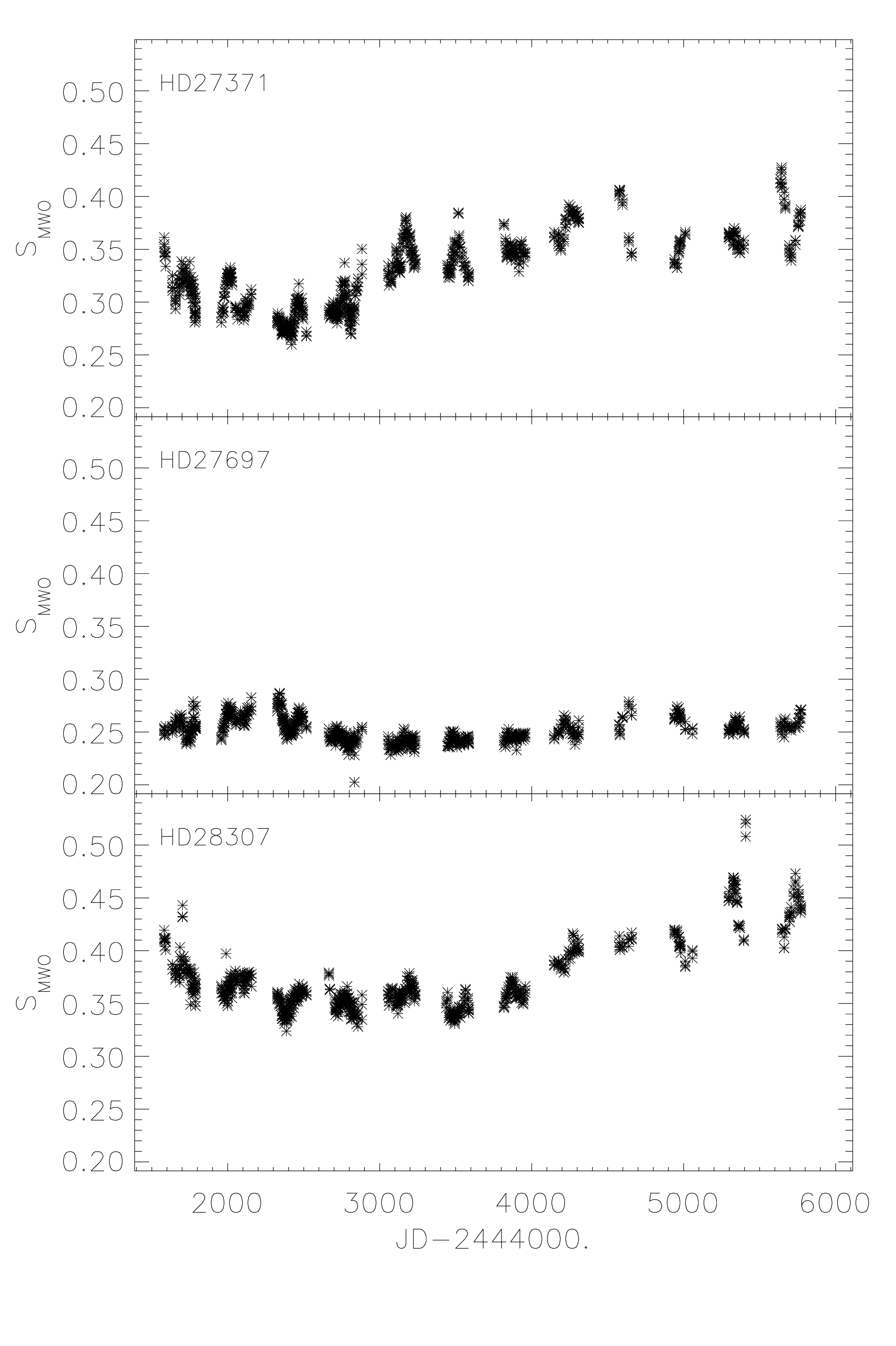}
	\end{tabular}
        \caption{Mt. Wilson S-index times series of the Hyades giants
          HD~27371 (top panel),
  	  HD~27697 (medium panel) and  HD~28307 (bottom panel).
         The zero point corresponds
	 to Jan 30, 1982 and data have been obtained in twelve
         consecutive observing seasons; see text for details.} 
	\label{fig_mtwilson}
\end{figure}

\subsection{Chromospheric activity}

\subsubsection{Ca II H\&K emission: the S-index}

For this paper we use the results of spectral monitoring 
with our TIGRE facility \citep{schmitt2014}, as well as the
S-index time series obtained in the framework of the Mount
Wilson H\&K project.  The TIGRE facility is a 
fully robotic telescope with a 1.2~m aperture, located at the La Luz
Observatory near Guanajuato, Mexico. Its only instrument is the two channel 
fiber-fed \'Echelle spectrograph HEROS with the wavelength range 
from 3800 \AA~to 8800 \AA~with a 100 \AA~ gap at 5800 \AA~ and a spectral
resolution of, according to our recent performance measurements,
R$\approx$21000; a detailed description of TIGRE is given by \cite{schmitt2014}.

In addition we use data obtained by the Mount Wilson H\&K Project, which have
been made available to the public and can be downloaded 
from ftp://solis.nso.edu/MountWilson\char`_HK/; 
a detailed description of the data is also provided at this 
web site. The available data specifically include the star identification; 
the calibrated S index, which we use in this paper as the basis for our
analysis; 
a code indicating with which instrument the data was taken; 
as well as the date of the observation and other material.

It is important to keep in mind that the hardware used by the TIGRE and
Mount Wilson H\&K Projects is fundamentally different. While TIGRE is using
an \'Echelle spectrograph, which covers, in particular, the whole region of the
Ca~II H and K lines, the Mount Wilson H\&K Project used a four channel
photometer, where the four channels were realized by an exit multi-slit
configuration on a rotating disk,
which allowed subsequent measurements of the output channels with a frequency of
about 30~Hz; a detailed description of this hardware is provided by
\cite{vaughan1978}. In this fashion measurements of the so-called R-band
(between 3991.067~\AA \ and 4011.067~\AA) and V-band (between 3891.067~\AA \
and 3911.069~\AA) were obtained, while the actual H and K fluxes
were measured in two narrow bands with a FWHM of 1.09~\AA \,
centered on the H and K lines
respectively.  The S-index was then calculated from the relation
\begin{equation}
\label{s_index}
S = \alpha \frac{N_H+N_K}{N_R+N_V},
\end{equation}
where the quantity $N_i$ denote the recorded number of counts in the band $i$,
and $\alpha$ is a correction factor, which is used to make measurements with
different hardware compatible with each other.

This definition and simultaneous measurement procedure of the S-index provides 
an activity record which is largely independent of
atmospheric conditions, since changes in atmospheric throughput or
transmission cancel out by its definition as a ratio between two measurements.  
Since it also comes with a list of over 40 calibration stars, it can be used
independently of the instrumentation for comparison with measurements from
over six decades ago to look for long-term  variability of chromospheric
activity of any star already observed back then 
at Mt. Wilson (see \cite{schro2018} for a more detailed discussion). 
On the other hand, a lot of the spectral information is lost in the
S-index construction, which is of course retained in the TIGRE spectroscopic
data, yet it is very useful to compute an S-index from the TIGRE data with
a procedure described in detail by \cite{mittag2015}. 

In our TIGRE spectra, we can inspect the emission line profiles to check
for their width, which is wider than the emission of main sequence stars,
according to the Wilson-Bappu effect \citep{wilson1957}. The TIGRE
S-index is based on the well-calibrated 1 \AA\ bandpath, and luminous giant
Ca II H\&K emission exceeds that width. For all Hyades giants, however,
we find that always about 95\% of their chromospheric emission is included.
Consequently, 
its variations are well proportional to the variable surface flux 
caused by chromospheric heating of the K giants and have no other source
(i.e., we find no noticeable variations in wavelength relative to the 
photospheric profile).

\subsubsection{Mount Wilson observations}

An important aspect for studying long-term variability of stellar magnetic
activity is that S-index measures are available from over six decades 
ago (see \cite{schro2018} for a more detailed discussion), and the Mount 
Wilson data base includes time series measurements of
the Hyades K giants HD~28307, HD~27371 and HD~27697. Nevertheless, only the
data for HD~28307 seem to have been published (see Fig.~1 in \cite{choi1995}).
For HD~28305 only a single S-index measurement exists, but no time series.

As pointed out and described by \cite{choi1995}, those measurements
were taken with a 2~\AA\, wide bandpass dedicated to giants, to mitigate the
above-mentioned Wilson-Bappu effect, and are therefore not directly
comparable to measurements taken with the ``standard'' narrow bandpass,
as used by TIGRE S-measurements and all earlier Mount Wilson observations. 

In Fig.~\ref{fig_mtwilson} we show the Mount Wilson S-index data recorded in the
time between 1984 and 1992; note that the data for HD~28307 have already been
shown by \cite{choi1995}.  All data sets appear to suggest the presence of
activity-cycles with periods on the order of 15~years, which is the time
span of the observations, however, clearly, this time span of these data
is too short to ascertain the existence of cycles with certainty.  
In addition, the data also show short-term variations, which can be interpreted
as rotational modulation and which we will in detail discuss below.

Regardless of such
long-term variations, Fig.~\ref{fig_mtwilson} shows that the S-indices of the
two X-ray active giants HD~28307 and HD~27371 are much higher than that of the
X-ray weak giant HD~27697, thus the chromospheric emission appears to
mirror the X-ray emission well.

\subsubsection{TIGRE observations}

For reasons of simplicity and calibration, we here now discuss the Hyades
K giants activity levels in terms of the standard Mt. Wilson S-index
$S_{\rm MWO}$, obtained from TIGRE/HEROS blue channel spectra of the
years 2014-2019 in the well-calibrated 1 \AA\ bandpass.
 Ever since the start of the TIGRE robotic observations in 2014, spectra
of the four Hyades giants have been taken on a regular schedule.  

Averages and the observed ranges of $S_{\rm MWO}$ for the four Hyades K 
giants from 6 years of TIGRE monitoring (2014 to 2019) 
are as given in Table 1 and can be summarized as follows: 
HD 28305 varies between $S_{\rm MWO}$ = 0.120 and 0.150 with an average of 
$<S>=0.129$, HD 27697 between 0.120 and 0.155 with $<S>$ = 0.133, while
the less bright K giants HD 27371 and HD 28307 vary between 0.145 and 
0.205 with $<S>$ = 0.178, and 0.150 to 0.195, with $<S>$=0.166, 
respectively (see also Table 1). 

We should note, to put these values into
perspective, that a value of 0.12 corresponds to the chromospheric emission of 
giant stars known to be inactive (see respective Figs. in \cite{duncan1991} 
and Fig.~3 in \cite{schro2012}), i.e. with only a ``basal flux'' 
not related to stellar activity as such.

\begin{figure}
\centering
\begin{tabular}{c}
\includegraphics[width=0.90\linewidth,angle=-0]{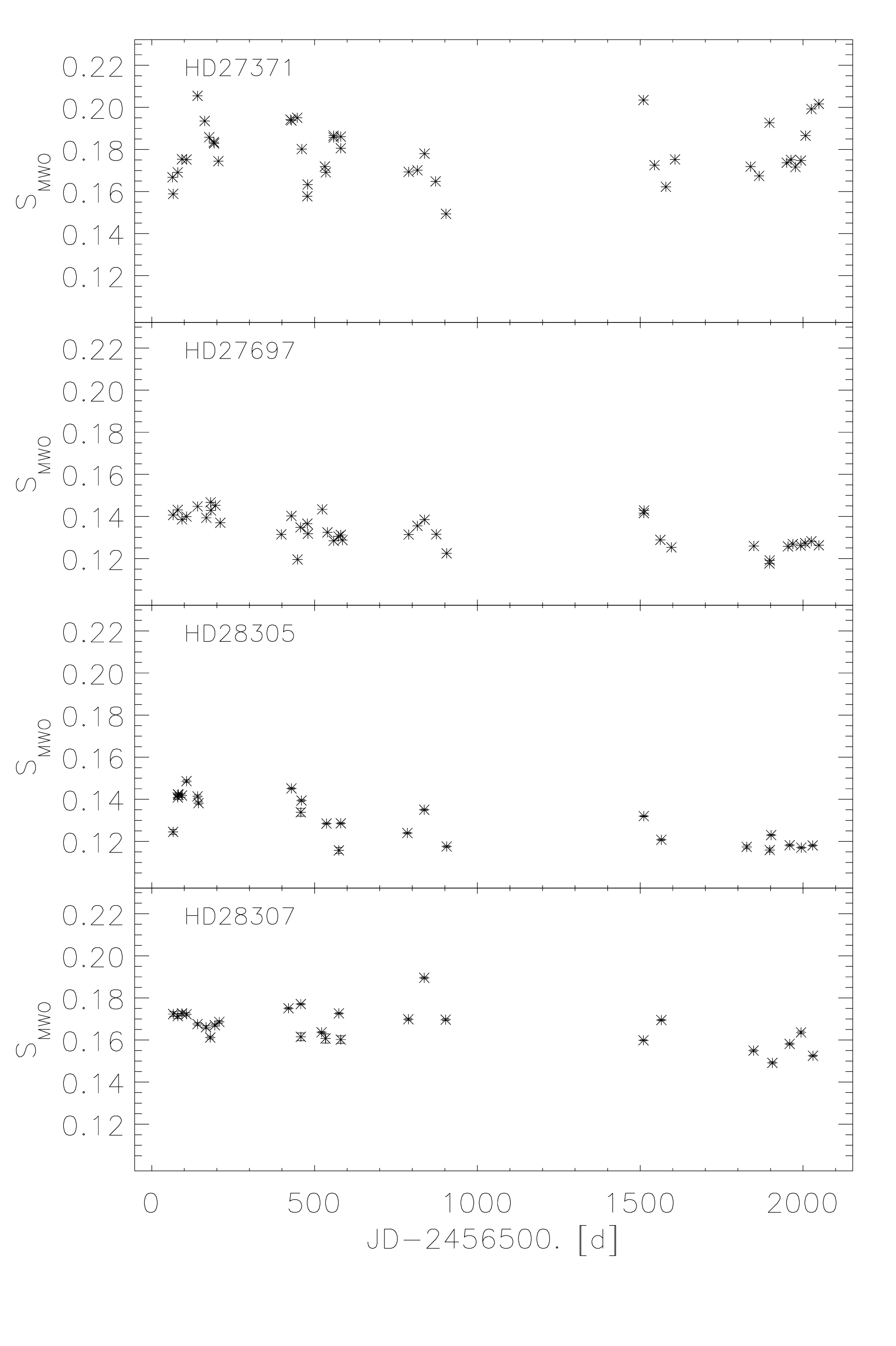}
\end{tabular}
\caption{TIGRE chromospheric monitoring of HD~28305 in the years of 2014
to 2019, using the calibrated S-index as defined by the Mt.~Wilson work
with 1 \AA\ H\&K line bandpass.}
\label{fig_tigre}
\end{figure}

If we take into account, that stars like the Sun have a 
somewhat larger ``basal flux'' level of S$_{\rm MWO}$ than giants, 
about 0.15 (see \cite{schro2012}), 
then the two active Hyades K giants resemble the activity level of 
the active Sun. Hence, on average, these are a bit more active than 
the Sun. Since these giants may be passing different phases of their 
cycles, an exact ranking between the two of them is premature.  

\begin{figure*}
\begin{minipage}{0.45\textwidth}
\begin{tabular}{c}
\includegraphics[width=0.95\linewidth,angle=-0]{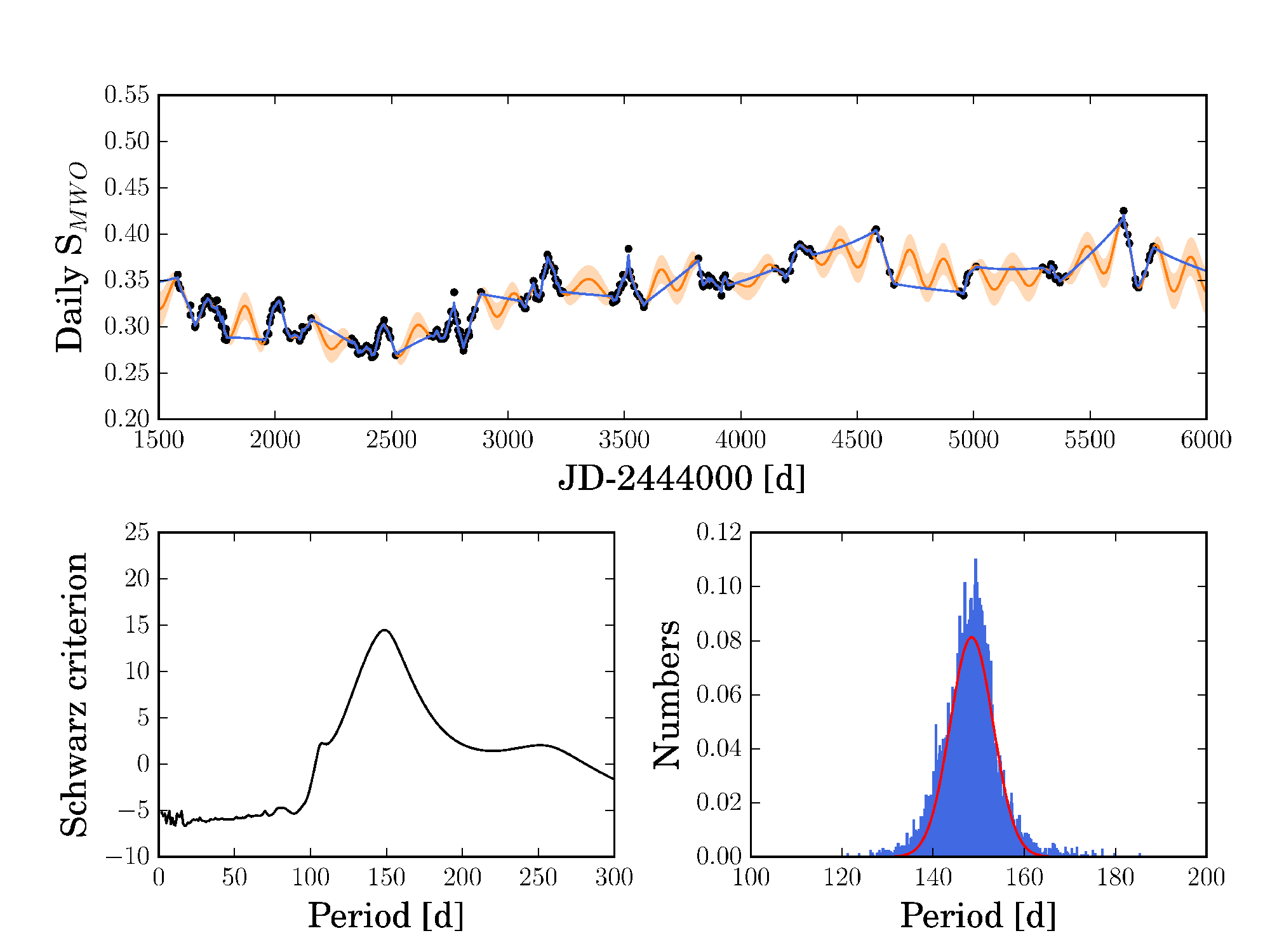}
\end{tabular}
\caption{Upper panel: Mount Wilson time series of HD~27371 (blue data points)
together with best fit GP (orange curve). Lower panel (left): Schwarz criterion
as function of trial period}.  Lower panel (right): Results of a Markov chain
Monte Carlo simulation of the trial periods; see text for details.
\label{gp1_lc}
\end{minipage}
\begin{minipage}{0.45\textwidth}
\begin{tabular}{c}
\includegraphics[width=0.95\linewidth,angle=-0]{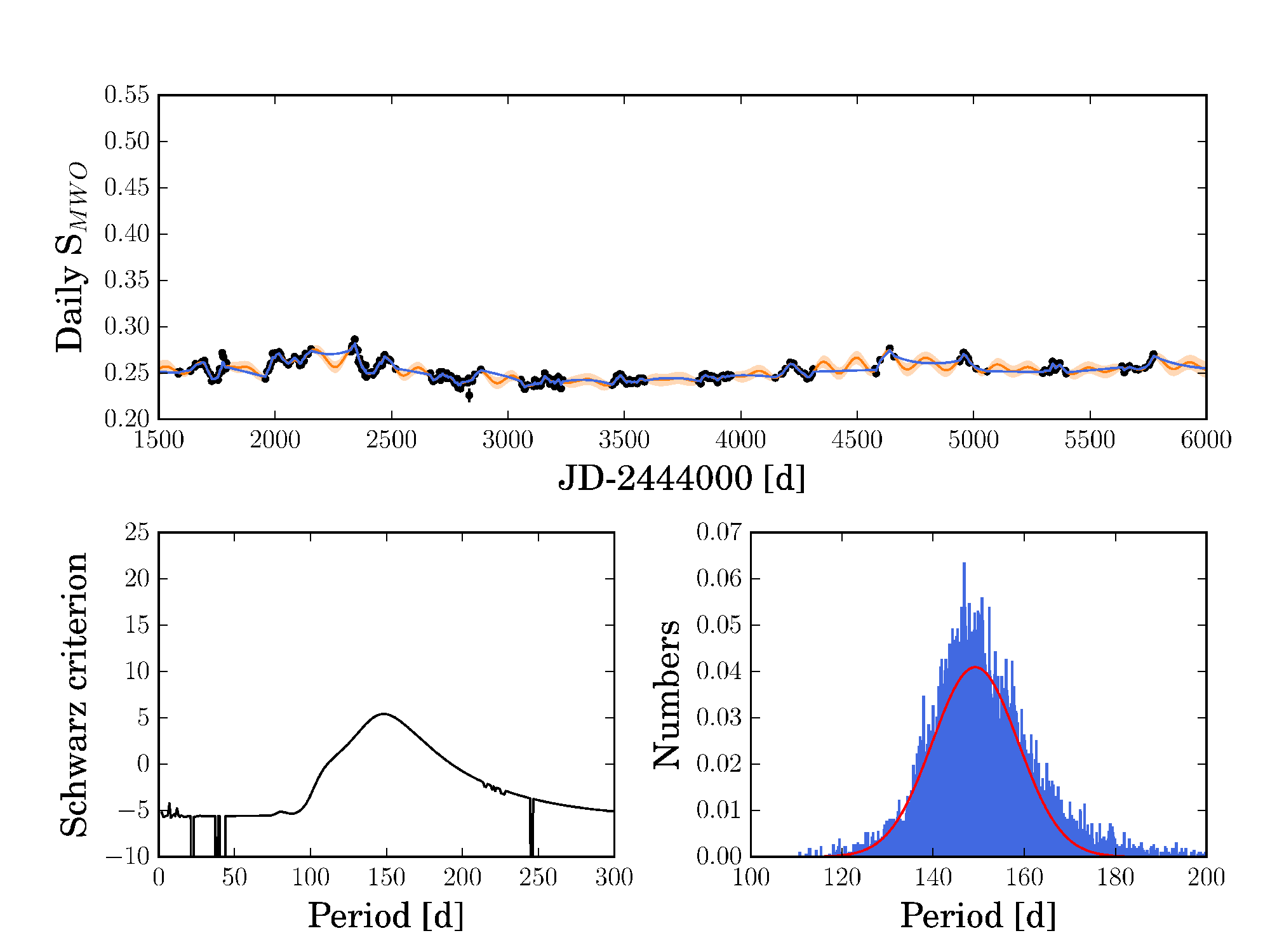}
\end{tabular}
\caption{Same as in Fig.~\ref{gp1_lc} with the Mount Wilson data for HD~27697.}
\label{gp2_lc}
\end{minipage}
\\
\begin{minipage}{0.45\textwidth}
\begin{tabular}{c}
\includegraphics[width=0.95\linewidth,angle=-0]{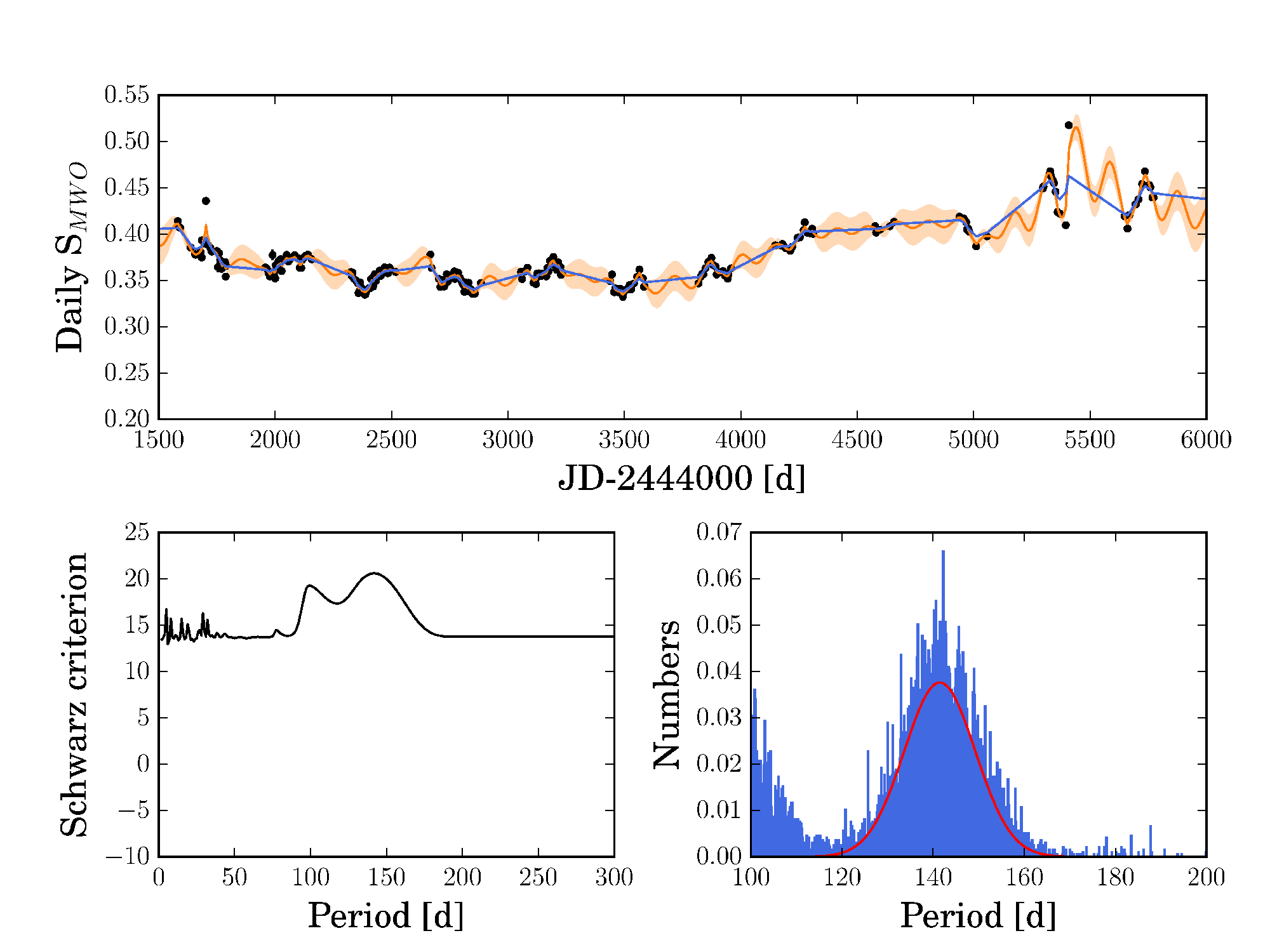}
\end{tabular}
\caption{Same as in Fig.~\ref{gp1_lc} with the Mount Wilson data for HD~28307.}
\label{gp3_lc}
\end{minipage}
\begin{minipage}{0.45\textwidth}
\begin{tabular}{c}
\includegraphics[width=0.95\linewidth,angle=-0]{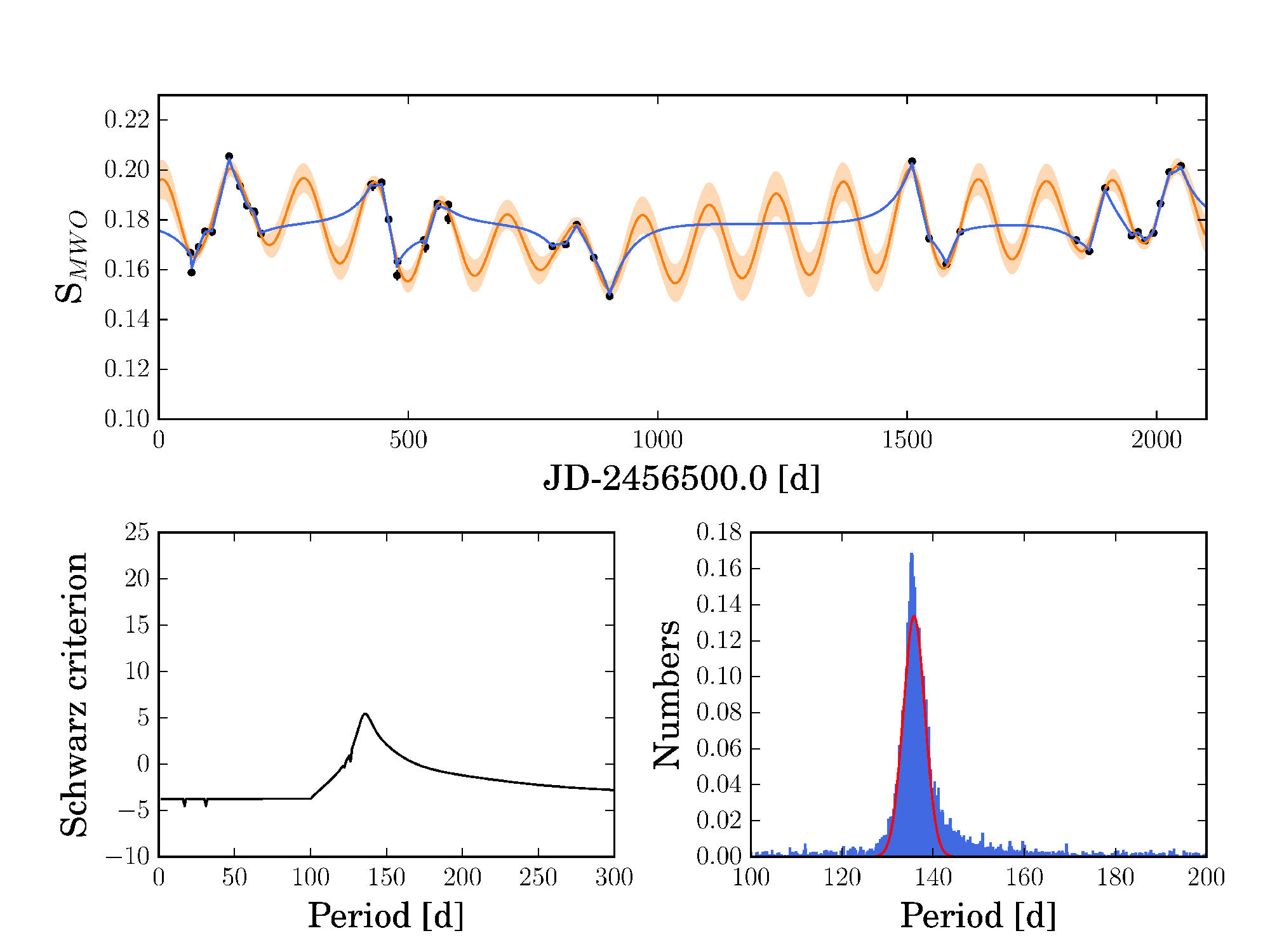}
\end{tabular}
\caption{Same as in Fig.~\ref{gp1_lc} with the TIGRE data for HD~27371.}
\label{gp4_lc}
\end{minipage}
\end{figure*}

\subsection{A clear relation between activity level and evolutionary state}

In general, the emerging 
picture is clear, both from coronal and chromospheric activity: 
The active pair of giants ($\gamma$ Tau = HD 27371 and 
$\theta^1$ Tau = HD 28307) is emitting X-ray surface fluxes  about two 
orders of magnitude larger than does the inactive pair. The latter 
($\epsilon$ Tau = HD 28305 and $\delta$ Tau = HD 27697) hardly reach the 
minimal coronal surface flux of inactive stars found 
by \cite{schmitt1997}, $F_{x,min} = o(10^4\,\mathrm{erg~cm^{-2} sec^{-1}})$,
which resembles $F_x$ of the inactive solar corona, when the star in 
question is on the left side of the ``corona-wind dividing line'' in the 
HRD. According to \cite{linshaisch1979}), coronal X-ray emission
is predominantly found on that left side, while the right (cool) side
in the HRD is dominated by cool winds; we note in this context,
that the Hyades K giants are very close to this ``dividing line''.

The same pattern, a split-down into two active and two inactive K giants, 
is reflected by the chromospheric activity, see the average S-values obtained
above for these giants (and see Table 1).  
It is now instructive to relate the different activity levels of the stars
$\gamma$ Tau (= HD~27371) and $\theta^1$ Tau (= HD~28307) on the one hand,
and $\epsilon$ Tau (= HD 28305) and $\delta$ Tau (= HD 27697) on the other,
with their positions on the evolutionary tracks
(shown in Fig.~\ref{fig_track}):

The two active giants $\theta ^1$ Tau $\gamma$ Tau), 
are only in the beginning of their central helium burning phase, implying that
they have just passed the fast RGB, and that their core contraction in the Hertzsprung gap 
is still very recent. By contrast, the stars $\epsilon$ Tau and $\delta$ Tau 
are already in the end stage of their central helium-burning, and have already remained 
in this stable phase for about 130~Myrs. Their much lower activity
thus suggests magnetic braking during this relatively stable period.
This process would then be comparable to those in main-sequence stars of
lower masses, which have convective envelopes and magnetic activity 
during their stable central hydrogen-burning phase,
i.e. stars with $M < 1.5 M_{\odot}$, significantly less massive than
the Hyades K giants.

\subsection{Rotation periods from S-data variability}

\begin{table}
\caption{Rotational periods}
\begin{small}
\begin{center}
\begin{tabular}{l | l |l  } 
\hline
 star & $P_{\rm rot}$ & S. criterion \\ 
      & [$d$]& \\
\hline
\multicolumn{3}{|c|}{Mount Wilson time series} \\
\hline
HD 27697/$\delta$ Tau    & 148.2$\pm$9.4 & 5.4 \\
HD 27371/$\gamma$ Tau    & 148.8$\pm$4.7 & 14.5  \\
HD 28307/$\theta^1$ Tau  & 141.9$\pm$7.8 & 20.6 \\
\hline
\multicolumn{3}{|c|}{TIGRE time series} \\
\hline
HD 27371/$\gamma$ Tau    & 136.0$\pm$2.4 & 5.5 \\
\hline
\end{tabular}
\end{center}
\end{small}
\label{Tab:SynPar}
\end{table}

At first glance, one might expect, just like in the case of main sequence
stars, that rotation periods can be obtained from activity monitoring for
giant stars as well -- utilizing the fact that active regions 
can be unevenly distributed on the stellar surface.

Nevertheless, the much longer rotation periods of giants make a difference: 
First, a sufficiently longer monitoring 
time span is required. While this is a merely technical problem, 
the limited lifetime of active regions, apparently of the same order of, or
shorter than the giant rotation periods, produces a serious 
interference with the search for the rotational signal. The appearances and 
disappearances of individual active regions 
result in peaks in any periodogram, which compete with and 
confuse the rotational signal. 
This problem is even larger for very inactive giants like HD 28305,
which have no or only short-lived and small active regions. 
But even very active stars do not always give good results, since 
the observer has to wait until an uneven distribution of active regions 
occurs. During 2014 to 2019, TIGRE data of HD 28307 ($\theta^1$ Tau)
turned out to be of little use because of a lack of a necessarily 
uneven activity region distribution.

In their study of M dwarfs, \cite{fuhrmeister2019} compare different period 
search algorithms and conclude that Gaussian Process (GP) modeling leads to 
the smallest number of false detections.  A clear advantage of GP
modeling is the fact that because of the stochastic nature phase shifts can 
be easily accommodated, while Fourier-based methods become more cumbersome.
Hence, we here settled for this approach. Nevertheless, considering the
aforesaid, we see the here presented rotation periods as preliminary.

In several Hyades giants' S-index time series, variations are clearly visible
already to the educated eye, see Fig.~\ref{fig_mtwilson} and
Fig.~\ref{fig_tigre}. 
Their GP analysis can be thought of as sets of random variables, of 
which any finite set has a joint normal distribution (see the book by 
\cite{rasmussen2006} for a detailed discussion of GPs).  
Hence, a GP is completely specified by a mean and a 
co-variance function, and the latter can be used to search for
periodic variations.   The chosen kernel function, which describes
the covariance of the data, forms the basis of all GP modeling. 

Here we use the kernel of the so-called {\it celerite} 
approximation \cite{foreman2017}, which takes the form
\begin{equation}
  k_{ij}  = \frac{a}{2~+~b} e^{-c \tau_{ij}} (cos(\frac{2 \pi \tau_{ij}}{P}) + 1 + b)
  + \delta_{ij} \sigma^2,
\label{celerite}
\end{equation}
where $k_{ij}$ is the co-variance matrix, $\tau_{ij}$ is the modulus of the 
time difference of the time stamps $t_i$ and $t_j$,
$a$ and $b$ are normalization constants, $c$ describes the lifetime of the 
features, $P$ is the desired period and $\sigma$ a term generating white noise;
we refer to \cite{fuhrmeister2019} for a more detailed description and
discussion 
of the applied procedure. The kernel in the form of Eq.~\ref{celerite} has the 
property that the kernel matrix $K$ can be inverted with O($n log(n)$)
operations rather than by O($n^2$). This provides a massive advantage
for large data sets, since the co-variance matrix of 
the data needs to be inverted many times during the modeling process.

The results of our GP modeling of the Mt. Wilson data are shown in 
Figs~\ref{gp1_lc}  for HD~28307, in Fig.~\ref{gp2_lc}  for
HD~27697, and nin Fig.~\ref{gp3_lc}  for
HD~27371.   Since the chosen {\it python} implementation of {\it celerite} 
can handle only
a single periodicity, we rectified the data by taking out the long-term 
cyclic variations, thereby concentrating on shorter term variability in 
which we expect to find the rotational signal.  

In each of the GP diagrams
(i.e., Figs.~\ref{gp1_lc},\ref{gp2_lc},\ref{gp3_lc}) we show the rectified 
time-dependence (black data points), the best-fit (in a maximum likelihood 
sense) model (purple solid line) as well as the model ``error'', i.e., the 
light purple shaded regions; clearly this ``error'' becomes
largest during those times when no data is available.   To better assess 
the modeling parameters we performed a Markov Chain Monte Carlo Analysis 
(MCMC) using the Python implementation of {\it emcee} \cite{foreman2013}, 
and present (in Figs.~\ref{gp1_lc},\ref{gp2_lc},\ref{gp3_lc})
the resulting period-likelihood scatter plots from runs with 24000 MCMC 
realizations These figures show well defined peaks and hence preferred 
periods of the Mount Wilson data, which we estimate (from the
``classical'' $\Delta$ likelihood $\le$ 1 approach as) to be
$P_{HD~27697}$ $\approx$ 148 $\pm$ 9~days,
$P_{HD~27371}$ $\approx$ 149 $\pm$ 5~days,
and $P_{HD~28307}$ $\approx$ 142 $\pm$ 8~days,  see Table 2.

The so far six years of TIGRE data still cannot compete with the 
time-coverage of the Mount Wilson 
data, but the very good s/n of our HEROS spectrograph cameras should 
produce S-index series 
and calibration, which better resolve smaller physical variations and 
introduce less noise to the analysis. So far, TIGRE S-data yield a 
credible result already for one star:

For the active K giant, HD 27371 ($\gamma$ Tau), 
TIGRE S-data produce a significant period of 137 days, which agrees, 
within the uncertainties, with the above value obtained from the Mount Wilson 
data (149 days, see Table 2). This is remarkable, given the independence, by 
instrumentation and epoch, of the respective data sets. Hence, for 
this star we are confident of a rotation period of about 143 days, when
combining both datasets.

We also note that \cite{auriere2015}, based on \cite{choi1995}, 
already quote a rotation period of 
140 days for the other active K giant HD~28307, using the very 
same Mount Wilson data, and in excellent agreement with our own analysis 
(142 days). 

Furthermore, given the well-defined and very suggestive
variations in the S-data series of both active Hyades giants, 
confidence in both these 142 day rotation periods is well justified.

However, the case of the inactive K giant 
HD~27697 is very different, its variations are very small. To us, 
therefore, it is not clear, whether its variation timescale of 
148 days is a physical variation at all, and if so,
really indicates the rotation period of that star. See below for a
further argument, why we doubt that the hardly active
giant HD~27697 should rotate as fast as its two active peers.

\subsection{Estimate of the convective turnover-time}

Inspecting the chromospheric and coronal activity of the Hyades giants
as given in \ref{Tab:SynPar}, we here showed a remarkably clear decrease
in the average S-value and X-ray luminosity over the
duration of central helium-burning of the Hyades K giants, much like 
we know it since long from cool main-sequence stars.

Unfortunately, there is no direct evidence of rotational spin-down
for giant stars: For one of the two inactive giants (HD~28305) we have no
rotation period at all, and the period obtained for the other one,
HD~27697, appears to be of a lesser significance and is similar
to the rotation periods of the active giants. 

However, the very active giants HD~27371 and HD28307, which are just
starting central helium burning, have trustable rotation periods of about
143 days. Hence, their rotation is five times slower than the solar rotation,
despite their larger activity. Consequently, since their slower rotation
cannot be an indicator of lower activity, it rather seems to relate to the very
different structure of a K giant compared to the Sun.

In mean-field dynamo theory, the decisive parameters is the 
Rossby number, i.e., the ratio between rotational period and convective 
turn-over time of the given stellar structure. Depending on a suitable 
definition of the latter (see discussion and references in
\cite{mittag2018}) mean-field dynamo activity seems to work only for Rossby
numbers smaller than unity.  Whether or not this also holds for giants is unclear, 
but for the following discussion we assume this to be the case. 
Then, in the non-active limit,
the rotational period becomes an empirical measure of the convective
turn-over timescale and consequently, rotation periods of giants, and how
they differ from convective-envelope main sequence stars, carry 
information on the giants' convective layers, where their dynamo is in operation.

Based on this idea, we estimate the empirical convective turn-over
timescale of a K giant from scaling the solar case by the factor, by which
the rotation period is longer than of a solar-type star of similar activity
level, as studied empirically by \cite{mittag2018}. The two active Hyades
K giants, with  rotation periods of around 143 days, have an activity level,
which compares to solar-like stars with rotation periods on the order of
15 days (i.e. stars more active than the Sun), suggesting a scaling-factor
in the rotation periods (K giants versus solar type stars) of about 10.  

As shown by \cite{mittag2018}, the 
upper envelope to the observed stellar rotation period distribution
coincides with stars of vanishing activity and a Rossby-number of
unity, and so gives a good empirical estimate of the convective turn-over time.
For stars like the Sun  that work puts this empirical value at
about 35.5 days (see 4$^{th}$ line in their Table 1), consistent with
rotating stellar evolution models of \cite{kimdem1996}, which follow in
detail both meridional mixing and convection and obtain a ``global''
(or nonlocal) convective turn-over time for their
solar model of 37.5 days. That value is the total travel-time of an
imaginary bubble for rising through the whole convection zone.

Other studies use a ``local''
convection turn-over time for, mainly, half a pressure scale height $H_P$ above the
bottom of the convective layer. In the case of the Sun, classical dynamo
models expect there the creation of the longitudinal magnetic field.
The local convective turn-over time represents the average (local) travel-time
of a bubble rising up by one convection length  $l_c = \alpha_c \cdot H_P$.
Hence, such different definitions and details of how the convection is
described, result in different absolute scales of the respective
computed convection turn-over times.
Individual values obtained from different codes and empirical timescales
may therefore differ from each other by at least a factor of two.  

If we now apply the same rotation period scaling-factor from above,
of about 10, to the minimal rotation period of the inactive Hyades K giants,
we find values of about 350 days for  HD~28305 and HD~27697. 
As discussed in the previous section, it seems impossible to verify this
estimate empirically, unless we are lucky enough to observe an exceptional, 
long-lasting active region on one of these rather inactive giant stars.

Since this suggested 350 days minimal rotation period
should characterize the convective turn-over timescale of the Hyades
K giants, we can consult our stellar models to discuss the factor
between the solar and the giant value -- at least on a relative scale,
as pointed out above. For this approach, the local convective turn-over
time is a very practical quantity, as it is easy to retrieve from any
stellar model, which is using the classical mixing-length theory.
In this approach the average convection velocity at a radial point $r$ is given 
by the expression

$v_c(r) = \sqrt{\alpha H_P g(r) \Delta{T(r)} / 2 T(r)}$

\hspace{2cm} $ = \sqrt{P(r) \Delta{T(r)}/\rho(r) T(r)}$

\noindent where our code specifies at each of its 200 height points
and at each time step the adiabatic bubble's temperature gain $\Delta{T(r)}$,
the global gas properties $T(r)$, $P(r)$, and $\rho$, using $\alpha=2.0$.
The local convective turn-over time is then given by
$\tau_c = \alpha H_P / v_c$, considering that $H_P = P / g \rho$ and that all
these values are here taken for a radial point at half a
pressure scale height above the bottom of the convection zone;
note that only half a pressure scale height higher up, the velocities
already increase by a factor of 2, which makes the absolute scale of
local convective turn-over times very dependent on where they are taken.

On a relative scale, though, the picture emerging from our stellar
models is quite instructive. Our solar model yields a local
convective turn-over time of 17 days. Considering that a global
turn-over time must be longer by some considerable factor, this is
in good agreement with \cite{kimdem1996}, despite using here
a much less sophisticated code. However, with our robust and fast
code, we can compute stellar models far beyond the main sequence.
A model matching the Hyades giants then suggests 450 days of
local convection turn-over time.

Note, that the surface gravity of the K giant model decreases from the solar
value by two orders of magnitude. Consequently, at the bottom of its large
convection zone, the main differences occur in the pressure scale height,
which rises by a factor of 20 in the K giant model, when compared to the solar
model. Hence, the much larger pressure scale heights of the giant's convective
envelope are the main driver for an almost 30 times larger convective turn-over time
as compared to solar-type stars. However, the empirically expected factor
(by how the rotation periods scale) is of only a factor 10, scaling
nearly with $g^{-1}$.

The discrepancy between empirical timescales and what our models suggest
does not much differ from earlier computational work by \cite{auriere2015}
(see their Fig.~5, and \cite{char2017} for a more detailed description):
Their local convective turn-over times near the bottom of the convection
zone become very large for luminous red giants, as they seem to scale
approximately with $g^{-2}$. 

However, there is a fundamental problem with operating a solar-type
dynamo in a giant, and this is probably the key to a better understanding:
It seems impossible to have magnetic field loops rise all the way 
through the huge convective envelope of a giant without having them
decay before reaching the photosphere, \citep{holzSchue2001}.  
And indeed, the discrepancies between empirical and model convective 
turn-over times in fact disappear, when we consider higher layers 
of the convective envelope of a giant star, where convective 
turnover-times are generally longer:
In our K giant models, the convection velocity increases
(and the local convective turn-over time decreases), driven by larger
temperature gains of the adiabatically rising bubbles, by 
up to an order of magnitude towards intermediate convection layers. 
This is more than enough to reconcile the lower empirical
factor of how much slower (only 10 times, not 30 times) the 
giant convection appears to be in the field-forming layer, 
when compared with the Sun.

Hence, giant convective envelopes may actually run  
a stratified dynamo (see \cite{brdbg2005}), if the 
milder gravity-scaling of empirical convective turn-over times
as suggested here is confirmed as a general behaviour
-- meaning that perhaps, with falling surface gravity, gradually
higher convective layers become involved in growing the 
longitudinal field in ever larger convective envelopes. 

\section{Discussion and conclusions}

The main result presented in this study is that the two K giants of 
2.75 $M_{\odot}$ ending central helium 
burning show much less chromospheric and coronal activity than 
the two of 2.62 $M_{\odot}$ beginning this 130 Myrs lasting phase. 
Even though we are dealing with only four stars, these observations are 
entirely consistent with the idea that magnetic activity, much like 
during central hydrogen burning
of cool ($M<1.5 M_{\odot}$, convective envelope) main sequence
stars, is decreasing during central helium burning of K giants, 
by the action of magnetic braking in both these cases. 

We therefore conclude that, first, despite 
the relaxation of the core with the onset of central helium burning, 
K giant clump chromospheres are heated by noticeable magnetic activity, 
which is consistent with the detection of coronal X-rays of such stars 
in the solar neighborhood, see e.g. \cite{huensch1996a}, and second, 
that magnetic braking dominates the evolution of magnetic activity 
during the stable phase of central helium burning.

We suggest magnetic braking to be at work during this phase.
By the relatively faster decrease of activity during central
helium-burning, when compared to central hydrogen-burning, 
it would be more than an order of magnitude faster than for cool 
main-sequence stars. That should in fact be expected, given that
a Hyades K giant is, by radius, over 15 times larger than a cool
main sequence star and so provides a much larger lever for magnetic 
coupling with its circumstellar and interstellar medium.  

These conclusions complement an emerging wider picture of angular momentum
evolution across the HRD: Asteroseismology work by \cite{beck2012} and, e.g. 
\cite{buys2016}, based on {\it Kepler} precision photometric monitoring,
has proven that the still contracting cores of giants on the foot of
the RGB are rotating up to ten times faster than the surface layers.
Apparently, fast core contraction of stars with $M > 1.5 M_{\odot}$
in the Hertzsprung gap drives a core spin-up, since X-ray detections
show them to have a strong activity
(see \cite{huensch1996a}, \cite{huensch1996b} and references given therein).

While it is not clear, how such a core spin-up could drive a dynamo and
of which type, by analogue, a similar process seems to work in massive stars
during their fast core contraction and ascent to the AGB, see the discussion 
by \cite{schro2018}: The detections of magnetic field and chromospheric
activity high up on the AGB demonstrate that, where in the HRD core contraction
is fast (as for fast evolving massive stars), activity increases.   

When, by contrast, the stellar core is fairly stable, 
as with K giants in central helium-burning, then magnetic braking 
seems to dominate the outcome. Furthermore, the resulting 
similarities between the activity of K giants and cool solar-type
stars raise the surprising question, whether we see the same type
of dynamo at work, despite their large structural differences. 

Unfortunately, surface gravities of AGB stars are so low that
the relevant pulsations to detect fast-spinning cores
of such stars are too slow, and resonances too broad, to be 
observable. Consequently, asteroseismology is unable to provide direct 
evidence for what happens inside AGB stars. We need to study them
by means of their rotational periods.

Consequently, further monitoring of chromospheric activity is required to
strengthen the still sparse observational evidence for the Hyades K
giants and other giants rotation periods. We hope to present 
such work in the near future, to derive approximate Rossby numbers 
from each activity degree and rotation period, in order to estimate 
empirical convective turnover-times for different giants and 
gravities. That information should provide important clues as to 
where in the giant convective layers its dynamo is actually working, and 
our studies of the four Hyades K giants already provide a 
first important milestone.

\section*{Acknowledgments}

This study made use of the services of the Strasbourg astronomical 
data centre. The authors are grateful for financial support by 
the joint bilateral project CONACyT-DFG No. 278156 and funding from 
the Deutsche Forschungsgemeinschaft in several related projects. 
We also wish to acknowledge the use of S-index data from 
the Mount Wilson Observatory HK Project, which was supported by both public 
and private funds through the Carnegie Observatories, the Mount Wilson 
Institute, and the Harvard-Smithsonian Center for Astrophysics starting in 
1966 and continuing for over 36 years.  These data are the result of the
dedicated work of O. Wilson, A. Vaughan, G. Preston, D. Duncan, S. Baliunas,
and many others. We further wish to acknowledge the spectra analysis tool
iSpec by Blanco-Cuaresma et al. and we want to thank the anonymous referee
for very helpful and concrete suggestions to improve this paper.

\bibliographystyle{mnras}

\label{lastpage}

\end{document}